\input epsf

\magnification\magstephalf
\overfullrule 0pt
\def\gsim{\raise.3ex\hbox{$\;>$\kern-.75em\lower1ex\hbox{$\sim$}$\;$}}

\font\rfont=cmr10 at 10 true pt
\def\ref#1{$^{\hbox{\rfont {[#1]}}}$}


\font\fourteenbf=cmbx12 scaled\magstep1

\font\tenbfit=cmbxti10
\font\sevenbfit=cmbxti10 at 7pt
\font\fivebfit=cmbxti10 at 5pt
\newfam\bfitfam 
\textfont\bfitfam=\tenbfit  \scriptfont\bfitfam=\sevenbfit
\scriptscriptfont\bfitfam=\fivebfit

\font\tenbfit=cmbxti10
\font\sevenbfit=cmbxti10 at 7pt
\font\fivebfit=cmbxti10 at 5pt
\newfam\bfitfam 
\textfont\bfitfam=\tenbfit  \scriptfont\bfitfam=\sevenbfit
\scriptscriptfont\bfitfam=\fivebfit

\font\tenbit=cmmib10
\newfam\bitfam
\textfont\bitfam=\tenbit%

\font\tenmbf=cmbx10
\font\sevenmbf=cmbx7
\font\fivembf=cmbx5
\newfam\mbffam
\textfont\mbffam=\tenmbf \scriptfont\mbffam=\sevenmbf
\scriptscriptfont\mbffam=\fivembf

\font\tenbsy=cmbsy10
\newfam\bsyfam 
\textfont\bsyfam=\tenbsy%

 \def\b {\beta}

\def\pmb#1{\setbox0=\hbox{#1}
 \kern.05em\copy0\kern-\wd0 \kern-.025em\raise.0433em\box0 }

\def\slash{/\kern-.5em}

\def \half {{\textstyle {1 \over 2}}}

\def\b{$\bullet~$}

\def\boxit#1{\vbox{\hrule\hbox{\vrule\kern1pt\vbox
{\kern1pt#1\kern1pt}\kern1pt\vrule}\hrule}}

\parskip=6pt
\parindent=0pt
\hsize=17truecm\hoffset=-5truemm
\vsize=23truecm
\def\footnoterule{\kern-3pt
\hrule width 17truecm \kern 2.6pt}


\catcode`\@=11 

\def\nolabels{\def\wrlabeL##1{}\def\eqlabeL##1{}\def\reflabeL##1{}}
\def\writelabels{\def\wrlabeL##1{\leavevmode\vadjust{\rlap{\smash%
{\line{{\escapechar=` \hfill\rlap{\sevenrm\hskip.03in\string##1}}}}}}}%
\def\eqlabeL##1{{\escapechar-1\rlap{\sevenrm\hskip.05in\string##1}}}%
\def\reflabeL##1{\noexpand\llap{\noexpand\sevenrm\string\string\string##1}}}
\nolabels
\global\newcount\refno \global\refno=1
\newwrite\rfile
\def\defref{$^{{\hbox{\rfont [\the\refno]}}}$\nref}
\def\nref#1{\xdef#1{\the\refno}\writedef{#1\leftbracket#1}%
\ifnum\refno=1\immediate\openout\rfile=refs.tmp\fi
\global\advance\refno by1\chardef\wfile=\rfile\immediate
\write\rfile{\noexpand\item{#1\ }\reflabeL{#1\hskip.31in}\pctsign}\findarg}
\def\findarg#1#{\begingroup\obeylines\newlinechar=`\^^M\pass@rg}
{\obeylines\gdef\pass@rg#1{\writ@line\relax #1^^M\hbox{}^^M}%
\gdef\writ@line#1^^M{\expandafter\toks0\expandafter{\striprel@x #1}%
\edef\next{\the\toks0}\ifx\next\em@rk\let\next=\endgroup\else\ifx\next\empty%
\else\immediate\write\wfile{\the\toks0}\fi\let\next=\writ@line\fi\next\relax}}
\def\striprel@x#1{} \def\em@rk{\hbox{}} 
\def\lref{\begingroup\obeylines\lr@f}
\def\lr@f#1#2{\gdef#1{\defref#1{#2}}\endgroup\unskip}
\newwrite\lfile
{\escapechar-1\xdef\pctsign{\string\%}\xdef\leftbracket{\string\{}
\xdef\rightbracket{\string\}}}

\def\writestop{\def\writestoppt{\immediate\write\lfile{\string\p
ageno%
\the\pageno\string\startrefs\leftbracket\the\refno\rightbracket%
\string\def\string\secsym\leftbracket\secsym\rightbracket%
\string\secno\the\secno\string\meqno\the\meqno}\immediate\closeout\lfile}}
\def\writestoppt{}\def\writedef#1{}
\catcode`\@=12 
\rightline{MAN/HEP/2011/18}
\rightline{DAMTP-2011-108}
\bigskip
\centerline{\fourteenbf ELASTIC SCATTERING AT THE LHC}
\vskip 8pt
\centerline{A Donnachie}
\centerline{School of Physics and Astromony, University of Manchester}
\vskip 5pt
\centerline{P V Landshoff}
\centerline{DAMTP, Cambridge University$^*$}
\footnote{}{$^*$ email addresses: Sandy.Donnachie@hep.manchester.ac.uk, \ 
pvl@damtp.cam.ac.uk}
\bigskip
{\bf Abstract}

The first data from the TOTEM experiment agree well with Regge theory, and 
demand a hard-pomeron contribution.
\bigskip\bigskip
{\bf 1 Introduction}

Invented half a century ago\defref\regge{
T Regge, Il Nuovo Cimento, 14 (1959) 951; G F Chew and S C Frautschi, 
Physical Review Letters 8 (1962) 41}, 
Regge theory is one of the major successes of
high energy physics. It describes high-energy scattering at small $t$ 
in terms of the exchanges of the known mesons and of other objects which 
are probably associated with glueball exchange. The latter are known as 
pomerons and the inclusion of one such exchange, that of the so-called soft 
pomeron, suffices to describe hadron-hadron scattering processes at small $t$.
\defref\book{
A Donnachie, H G Dosch, P V Landshoff and O Nachtmann, {\sl Pomeron Physics 
and QCD}, Cambridge University Press (2002)
}.

At large $t$ it is necessary to introduce an additional term into the $pp,\bar pp$ 
amplitudes, which we have identified
\defref\ggg{
A Donnachie and P V Landshoff, Physics Letters B387 (1996) 637
}
as arising from triple-gluon exchange.

The soft pomeron contributes to hadron-hadron total cross sections a term with
energy dependence
$$
s^{\epsilon_1}~~~~~\epsilon_1\approx 0.08
\eqno(1)
$$
Deep-inelastic lepton scattering data require the introduction of a second 
pomeron, the so-called hard pomeron 
\defref\hard{
A Donnachie and P V Landshoff, Physics Letters B437 (1998) 408
}. 
Although hard-pomeron exchange is not needed to describe hadron-hadron total 
cross sections up to energies $\sqrt s$ below 1~TeV, it may nevertheless be 
present, giving a small contribution with energy dependence
$$
s^{\epsilon_0}~~~~~\epsilon_0\approx 0.4
\eqno(2)
$$
We have pointed out
\defref\krakow{
P V Landshoff,  arXiv:0811.0260v1
}
that such a hard-pomeron contribution is certainly
present if the upper of the two contradictory measurements
\defref\tevatron{
CDF collaboration: F Abe et al, Physical Review D50 (1994) 5550;
E710 Collaboration, N.Amos et al: Physics Letters B243 (1990) 158
}
at the Tevatron should turn out to be correct, and that
this leads to a considerable uncertainty in the prediction for the total-cross
section at the LHC:
$$
\sigma^{\hbox{\sevenrm TOT}}=125 \pm 25 \hbox{mb}~~~~\sqrt{s}=14~\hbox{TeV}
\eqno(3)
$$
Without a hard pomeron contribution, the prediction is just over 100~mb at 
14~TeV, and close to 91~mb at 7~TeV. The TOTEM collaboration has now found
\defref\totemm{
TOTEM collaboration: G Antchev et al, Europhys Lett 96 (2011) 21002
}
$$
\sigma^{\hbox{\sevenrm TOT}}=98.3 \pm 0.2 \hbox{ (stat)} {\pm}2.8 \hbox{ 
(syst)}  \hbox{mb} ~~~~\sqrt{s}=7~\hbox{TeV}
\eqno(4)
$$

The TOTEM collaboration has also measured\ref\totemm
\defref\totem{
TOTEM collaboration: G Antchev et al, Europhys Lett  95 (2011) 41001
}
$pp$ elastic scattering. In this paper we use these data, together with
those for $pp$ and $\bar pp$ total and differential cross sections below 
1.8~TeV and for the proton structure function $F_2(x,Q^2)$ at small $x$ 
from HERA, to conclude that a moderate hard-pomeron contribution is indeed 
present in hadron-hadron scattering. The prediction (3) is now refined to
$$
\sigma^{\hbox{\sevenrm TOT}}=113 \pm 5 \hbox{mb}~~~~\sqrt{s}=14~\hbox{TeV}
\eqno(5)
$$ 

We note that there are alternative approaches combining soft and hard 
pomerons\defref\others{
A D Martin et al, arXiv:1011.0287;
E Gotsman et al, arXiv:0903.0247, arXiv:0901.1540
}.

\bigskip
{\bf 2 Data fit}

Some 20 years ago we fitted
\defref\sigtot{
A Donnachie and P V Landshoff, Physics Letters B296 (1992) 227
}
all existing $pp$ and $\bar pp$ total cross section data at energies
$\sqrt s$ greater than 10 GeV with a combination of two fixed powers of $s$, 
one corresponding to $\rho, \omega, f_2, a_2$ exchange and the other to
the soft-pomeron exchange of (1). 
We pointed out that these were to be regarded as effective powers, 
which were actually a combination of fixed powers arising from single exchanges
and more complicated terms from multiple exchanges. 
When the HERA data at small $x$ appeared it was immediately 
apparent \ref{\hard} that the same fixed powers of $1/x$ would fit the data 
only if another $(1/x)^{\epsilon_0}$ term with $\epsilon_0\approx 0.4$ were 
added. 

While previously we used the
separate data from the two HERA experiments, in this paper we use
the combined data
\defref\combined{
HERA data: www-h1.desy.de$/$psfiles$/$figures$/$d09-158.nce+p.txt
}. 

For lepton-induced reactions there are no constraints from unitarity, so
fixed powers of $1/x$ are allowed and we will continue to assume that they
represent a good approximation. But for hadron-hadron processes unitarity
is violated at large $\sqrt s$ if the fixed powers are not moderated by the
introduction of additional terms in the amplitude. 
The fixed powers result from the exchange of single particles, and
the additional terms correspond to multiple exchanges. Although
we know some general analytic properties of these multiple-exchange terms, 
a full numerical calculation of them is still beyond present knowledge.

One way to take account of these known analytic properties is to
write the amplitude as a Fourier transform over the two-dimensional momentum 
transfer {\bf q}, where ${{\bf q}}^2=-t$,
$$
A(s,t)={2is}\int d^2b~ e^{-i{\bf q}.{\bf b}}~\tilde A(s,b)
\eqno(6a)
$$
and write
$$
\log(1-\tilde A(s,b))=-\chi(s,b)
\eqno(6b)
$$
so that
$$
A(s,t)={2is}\int d^2b~ e^{-i{\bf q}.{\bf b}}\big (1-e^{-\chi(s,b)}\big )
={2is}\int d^2b~ e^{-i{\bf q}.{\bf b}}~\sum_{n=1}
{1\over n!}(-\chi(s,b))^n
\eqno(6c)
$$
This is known as the eikonal representation.

If we identify the $n=1$ term with the contribution from single exchanges 
$$
A_{\hbox{\sevenrm SINGLE}}(s,t)={2is}\int d^2b\, e^{-i{\bf q}.{\bf b}}~
\chi _S(s,b)
\eqno(7)
$$
then the further terms
have the correct analytic properties to describe the multiple exchanges. 
One possibility is to insert the function $\chi_S(s,b)$ that this gives into
the full expansion, but this is only a model and it has no theoretical
foundation. We prefer to
take as our model for double exchange 
$$
2is\lambda\int d^2b\, e^{-i{\bf q}.{\bf b}}~\half (\chi _S(s,b))^2
\eqno(8)
$$
where $\lambda$ is assumed to be a constant, whose value has to be fixed. 
We shall assume also that we can neglect the contributions from more
than double exchange ($n>2$). We stress that again this is only a model; as yet
we do not have the theoretical knowledge to go beyond this.

$pp$ elastic scattering data from the CERN ISR\defref\isr{
CHHAV collaboration: E Nagy et al, Nuclear Physics B150 (1979) 221
}
find a dip, which is deepest at $\sqrt s=30.54$ GeV at which energy it
is at $|t|=1.425$ GeV$^2$. 
We fix the value of $\lambda$ in (8) by requiring that the
imaginary part of the amplitude vanishes near the dip: it turns out that
$t=1.4$ GeV$^2$ is a good point to choose. 

The dip moves slowly inwards towards $t=0$ as $\sqrt s$ increases. This means 
that at values of $t$ on either side of the dip the amplitude varies 
rapidly with energy and so general principles require that it also have 
non-negligible real part. In order correctly to model the dip we have
to ensure that both the real and the imaginary parts become very small
at the same value of $t$. 

We introduce 4 Regge trajectories:
$$
\alpha_i(t)=1+\epsilon_i+\alpha_i't~~~~~
\eqno(9)
$$
with $i=0$ referring to the hard pomeron, $i=1$ to the soft pomeron, 
$i=2$ the degenerate trajectory for $f,a_2$ exchange and $i=3$ for
$\omega,\rho$ exchange. We fix $\alpha_1'$ at the value 0.25~Gev$^{-2}$ 
that has been known\defref\jaros{
G A Jaroskiewicz, Physical Review D10 (1974) 170
}
for nearly 40 years, and\defref\traj{
Reference [{\book}], figure 2.13
}
$\alpha_2'=0.8$ GeV$^{-2}, \alpha_3'=0.92$ GeV$^{-2}$. We try various values 
for $\alpha_0'$ and find that 0.1 GeV$^{-2}$ works well. We treat the
four values of the $\epsilon_i$ as free parameters.

For the high-energy $pp$ elastic amplitudes we use
$$
A(s,t)=\sum_{i=0}^3 Y_i~e^{-\half i\pi \alpha_i(t)}~(2\nu \alpha_i')^
{\alpha_i(t)} 
\eqno(10a)
$$
with
$$
2\nu=\half (s-u) ~~~~~ Y_i=-X_i~(i=0,1,2) ~~~~~ Y_3=iX_3
\eqno(10b)
$$
with $X_0,X_1,X_2,X_3$ real positive. The factor $i$ multiplying $X_3$ is
a manifestation of the Regge signature factor\ref{\book} for negative
$C$-parity exchange. The amplitude for $\bar pp$ scattering
is the same, except that $Y_3$ has the opposite sign. The normalisation of
the amplitudes is such that $\sigma^{\hbox{\sevenrm TOTAL}}=s^{-1}
$Im $A(s,t=0)$.

At large $t$ triple-gluon exchange contributes to the amplitude \ref{\ggg}
$$
Cst^{-4}
\eqno(11a)
$$
where the data give
$$
C=3.4~ \hbox{GeV}^{-4}
\eqno(11b)
$$
This form cannot be valid near $t=0$. We have tried various forms for small $t$
which match smoothly to (11a) at some value $t=t_0$, and find that
$$
{Cs\over t_0^4}~e^{4(1-t/t_0)}
\eqno(11c)
$$
works as well as any. Triple-gluon exchange, which is real, is important in
obtaining the correct dip structure. We adjust the value of $t_0$ so as to 
match the shape of the dip as best we can at $\sqrt{s} = 30.54$ GeV. We use 
$t_0=5.4$ GeV$^2$. 

For deep inelastic lepton scattering at small $x$
we use an expression which successfully fitted\ref\hard
the separate H1 and ZEUS data:
$$
F_2(x,Q^2)=\sum_{i=0}^2f_i(Q^2)(1/x)^{\epsilon_i}
\eqno(12a)
$$
with
$$
f_0(Q^2)=A_0{(Q^2/Q_0^2)^{1+\epsilon_0}\over(1+Q^2/Q_0^2)^{1+\epsilon_0 /2}}
$$$$
f_i(Q^2)=A_i\Big ({Q^2\over Q^2+Q_i^2}\Big )^{1+\epsilon_i}
~~~i=1,2
\eqno(12b)
$$
We have shown
\defref\evol{
A Donnachie and P V Landshoff, Physics Letters B533 (2002) 277
}
that this form for $f_0(Q^2)$ gives a variation with $Q^2$
that agrees to very high accuracy with DGLAP evolution, and have explained
that this is the only part of $F_2(x,Q^2)$ at small $x$ to which DGLAP
may validly be applied.

\topinsert
\centerline{\epsfxsize=0.48\hsize\epsfbox[75 600 300 760]{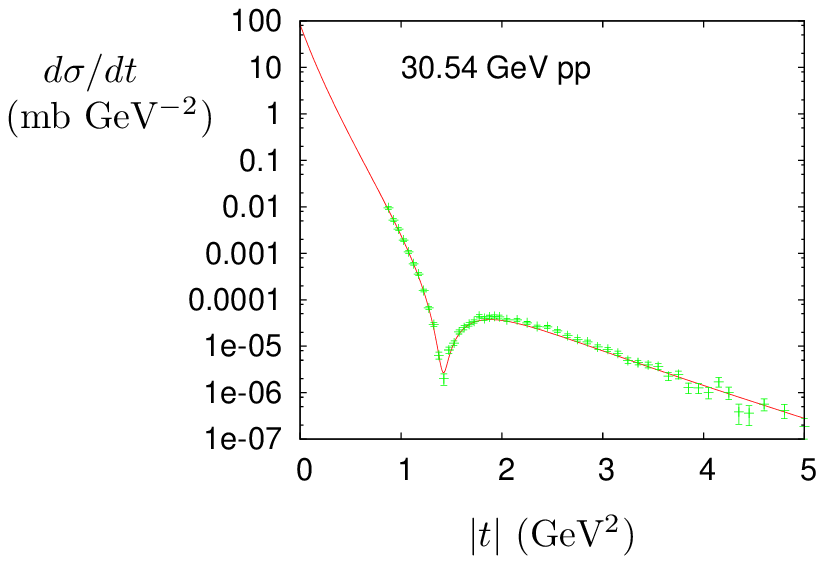}
\hfill\epsfxsize=0.48\hsize\epsfbox[75 600 300 760]{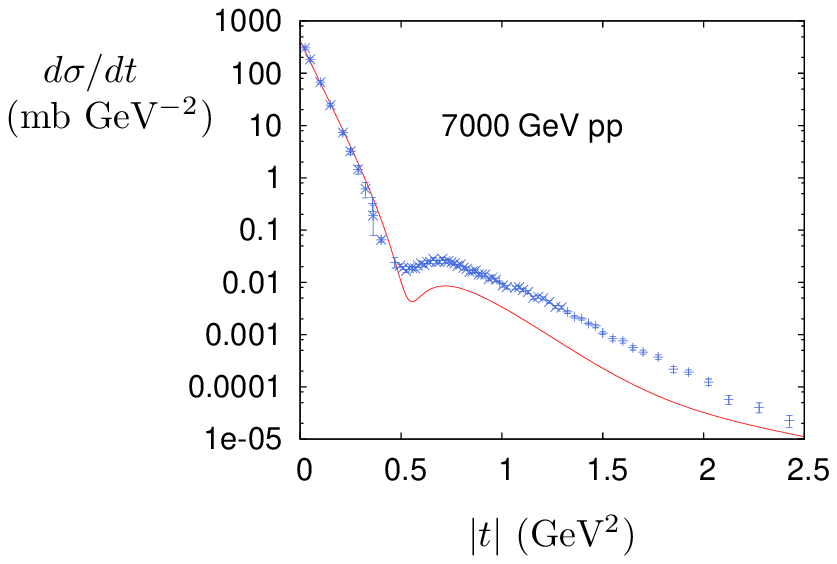}}

\centerline{Figure 1: 31 GeV and 7 TeV $pp$ elastic
scattering data with fit with no hard pomeron}
\endinsert

We use these forms for the $pp,\bar pp$ amplitudes and for $F_2(x,Q^2)$, 
including $Q^2=0$, to
make a simultaneous fit to the data for $F_2(x,Q^2)$ with $x<0.001$ and the
$pp,\bar pp$ 
total cross sections at energies above 10 GeV and below 1 TeV. Then we
adjust the value of $\lambda$ to correctly reproduce the dip in
the $pp$ elastic differential cross section at 30.54~GeV to obtain the 
fit shown in figure 1. 
As we have said, with no hard-pomeron term in the $pp$ amplitude this 
gives too low a cross section when extrapolated 
to 7 TeV. As is shown in figure 1, it also comes below the TOTEM elastic-scattering data. 

\topinsert
\centerline{\epsfxsize=0.48\hsize\epsfbox[75 600 300 760]{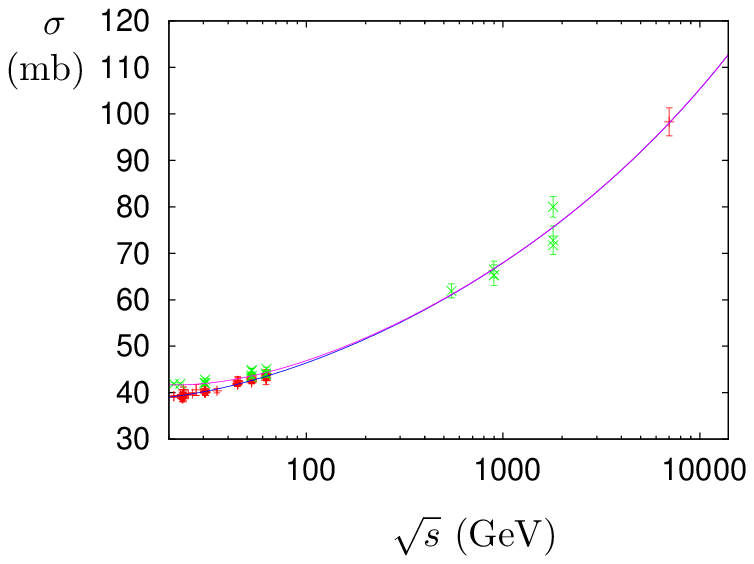}
\hfill\epsfxsize=0.48\hsize\epsfbox[75 600 300 760]{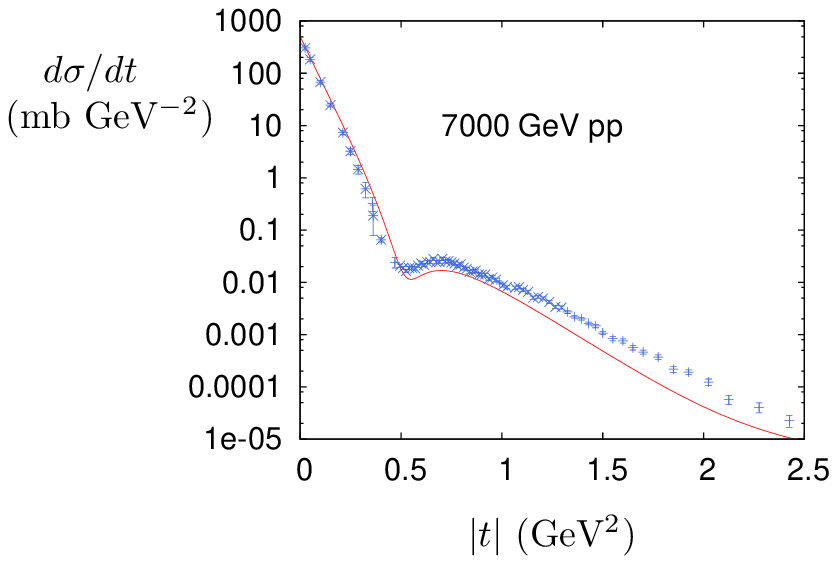}}

\centerline{Figure 2: $pp$ total cross section and elastic scattering
at 7 TeV with $X_0=1.2,\alpha_0'=0.1$}
\endinsert

In order to increase the total cross section so as to agree with the TOTEM 
measurement (4), we need to include a hard-pomeron contribution to the
$pp$ (and also $\bar pp$) total cross section. Figure 2 shows the fit
with $X_0=1.2$ and still $\alpha_0'=0.1$. This fit has
$$
A_0=0.063,~  A_1=0.315,~  A_2=0.229,~~  Q_0^2=2.95~\hbox{GeV },~  Q_1^2=794~
\hbox{MeV },  Q_3^2=303~\hbox{MeV}
$$$$
\epsilon_0=0.362,~  \epsilon_1=0.093,~ \epsilon_2=-0.360,~ \epsilon_3=-0.533~ 
$$$$
X0=1.2,~  X1= 243.5,~  X2= 246.9,~  X3= 136.7,~ \lambda=0.440
\eqno(13)
$$
\bigskip
{\bf 3 Discussion}

We make the following comments:

\b The combined HERA data favour a value of the hard-pomeron power $e_0$ 
some 10--20\% smaller than that we obtained from fitting the separate
H1 and ZEUS data.

\b The differences between the powers $\epsilon_2$ and $\epsilon_3$ 
corresponding to
the $C=+1$ and $C=-1$ particle exchanges is in accord with the powers
obtained from making a Chew-Frautschi plot\ref{\traj}. 

\b Our fit in figure 2 to the TOTEM elastic scattering data is not
perfect, but we only have a model: there is no known theory.  Our expression
for the double exchange has the correct analytic properties but its exact
form is not known, and it cannot be exactly correct to neglect the triple and
higher exchanges.

\b In our model, $\chi(s,b)$ in (6c) is taken to be given by
$$
1-e^{-\chi(s,b)}=\chi _S(s,b)-\half (\chi _S(s,b))^2
\eqno(14)
$$
Unitarity requires that Re $\chi(s,0)>0$, or
$$
|1-\chi _S(s,0)+\half (\chi _S(s,0))^2|<1
\eqno(15)
$$
We find that this is exceeded by 4.5\%, so again our model is not perfect.

\medskip\immediate\closeout\rfile\writestoppt
\baselineskip=14pt{{\bf References}}\bigskip{\frenchspacing%
\parindent=20pt\escapechar=` \input refs.tmp\bigskip}\nonfrenchspacing

\bye